\documentclass[review]{elsarticle}

\usepackage{lineno,hyperref}

\journal{Acta Astronautica}

\usepackage{amssymb}
\usepackage{latexsym}
\usepackage{siunitx}

\usepackage{url}
\usepackage[dvipsnames]{xcolor}

\definecolor{newcolor}{rgb}{.8,.349,.1}
\usepackage{atbegshi}

\newcommand{\hlreviewone}[1]{{#1}}
\newcommand{\hlreviewtwo}[1]{{#1}}
\newcommand{\hlreviewthree}[1]{{#1}}

\AtBeginDocument{\AtBeginShipoutNext{\AtBeginShipoutDiscard}}

\begin{document}

\begin{frontmatter}

\title{Sending a Spacecraft to Interstellar Comet 2I/Borisov}
\
\author[1]{Adam Hibberd\corref{cor1}}
\cortext[cor1]{Corresponding author}
\ead{adam.hibberd@i4is.org}
\author[1,2]{Nikolaos Perakis}
\author[1]{Andreas M. Hein}

\address[1]{Initiative for Interstellar Studies, 27-29 South Lambeth Road, London SW8 1SZ, United Kingdom}
\address[2]{Technical University of Munich, Boltzmannstr. 15, 85748 Garching, Germany}





\begin{abstract}
 \hlreviewone{On 2019 August 30}, a second interstellar object 2I/Borisov was discovered 2 years after the discovery of the first known interstellar object, 1I/'Oumuamua. Can we send a spacecraft to this object, using existing technologies? In this paper we assess the technical feasibility of a near-term mission to 2I/Borisov. We apply the Optimum Interplanetary Trajectory Software (OITS) tool to generate trajectories to 2I/Borisov. As results, we get the minimal $\Delta V$ trajectory with a launch date in 2018 July. For this trajectory, a Falcon Heavy launcher could have hauled an 8 ton spacecraft to 2I/Borisov. For a later launch date, results for a combined powered Jupiter flyby with a Solar Oberth maneuver are presented. For a launch in 2027, we could reach 2I/Borisov in 2052, using the Space Launch System (SLS), up-scaled Parker probe heat shield technology, and solid propulsion engines. Using a SLS a spacecraft with a mass of 765 \si{kg} could be sent to 2I/Borisov. A Falcon Heavy could deliver 202 \si{kg} to 2I/Borisov. Arrival times sooner than 2052 can potentially be achieved but with higher $\Delta V$ requirements and lower spacecraft payload masses. 2I/Borisov's discovery shortly after the discovery of 1I/'Oumuamua implies that the next interstellar object might be discovered in the near future. The feasibility of a mission to both, 1I/'Oumuamua and 2I/Borisov using \hlreviewthree{scaled versions of existing technologies} indicates that missions to at least some future interstellar objects are feasible as well. 
\end{abstract}

\end{frontmatter}


\section{Introduction}
\label{sec1}
On \hlreviewone{2017 October 19}, the first interstellar object (ISO) 1I/‘Oumuamua was detected \hlreviewone{\citep{2017Natur.552..378M,Knight2017,Fitzsimmons2018,Schneider2017,Feng2017,Wright2017}}. Almost two years later on \hlreviewone{2019 August 30} a second ISO, 2I/Borisov was discovered by Gennadiy Borisov using his MARGO observatory in Crimea. 
Unlike 1I/'Oumuamua, which remains somewhat of an enigma, 2I/Borisov was found to be a comet \citep{2019ApJ...886L..29J} similar to ones found in our solar system with an associated coma and tail as it approached and rounded the sun \citep{Bolin2020,DeLeon2019,Fitzsimmons2019,Guzik2020,Jewitt2020,2019ApJ...886L..29J,Manzini2020,Opitom2019,Xing2020,Ye2020}. Its apparent fragmentation around \hlreviewone{2020 February/March} (perihelion was on \hlreviewone{2019 December 8}) was later established to be an ejection of material, around 0.1\% of the total mass \citep{Jewitt_2020}. \\
Due to their origin, ISOs could provide unique insights into properties of other star systems and the interstellar environment \citep{eubanks2020exobodies}. Close-up observations of such objects would only be possible via spacecraft that perform a flyby or rendezvous with the ISO. We have previously shown that missions to 1I/'Oumuamua would be feasible using \hlreviewthree{scaled versions of} existing and near-term technologies and with launch dates from today to several decades into the future \citep{Hein2019,Hibberd2020}. \cite{Seligman2018} have shown that direct missions to ISOs are feasible on short notice. However, 1I/'Oumuamua and 2I/Borisov were both discovered too late to launch a spacecraft for direct intercept (i.e. without gravitational assists (GA)) which generally corresponds to the minimium $\Delta V$/maximum payload mass scenario for these type of objects. Hence, one or more GAs are needed to catch up with them. \hlreviewthree{Also of relevance here are \cite{Castillo2019} which investigates missions to long-period comets and interstellar objects (with GAs) and \cite{Vivan2020} which explores long-period comets from multiple staging orbits.} \\
To generate the high heliocentric hyperbolic excess speeds necessary to catch these fast-moving objects, a solar slingshot within a few Solar Radii of the sun (known as a Solar Oberth, SO, maneuver) is employed \citep{Hein2019,Hibberd2020}. A preceding $V_{\infty}$ Leveraging Maneuver is also utilised to reduce the $\Delta V$ needed to travel from Earth to Jupiter. Generally speaking some combination of GAs of the inner planets could also be exploited to enhance performance (i.e. generate greater payload masses to Jupiter), however this would depend on the specific orbit and timing of the ISO and relative alignment of the planets. Missions to 1I/'Oumuamua using Nuclear Thermal Rockets (NTR) have also been studied \citep{HIBBERD2021594} indicating a dramatic improvement in payload mass performance and journey time with respect to chemical. NTR has a lower Technical Readiness Level (TRL) of around 4 (NASA Technology Roadmap 2015) and so clearly more R\&D would be needed to realise this performance gain, consequently in this study chemical propulsion only is considered (note however the assumptions adopted also apply to NTR, though the payload masses would be considerably greater for NTR).\\
Thus, in this paper, we assess the principle feasibility of a mission to 2I/Borisov using \hlreviewthree{scaled versions of existing} technologies. \hlreviewthree{By 'existing', we neither mean 'off-the-shelf' nor TRL 9 (already flown on a similar system in a similar environment). For most of the technologies we consider here, 'existing' means that the basic technology exists but would need to be scaled up or combined in new ways, reducing its TRL to 5.}
\\
\hlreviewone{For the large encounter distances necessitated by the indirect trajectory elaborated in this paper ($>$ 200 au) it is pertinent to enquire what the scientific return would be. A mission would require exit out of the heliopause and into the Local Interstellar Medium which would alone enable serious and useful scientific observations and measurements. Various papers have considered s/c missions to explore the ISM, such as \cite{Brandt2019a,Friedman2014,Alkalai2017} or the outer solar system \citep{Turyshev2020}. Furthermore potential missions to ISOs have been examined for their potential scientific return in such papers as \cite{Hein2020a,Eubanks2020}, the latter of these papers analyses the spectroscopic return which would be gleaned by releasing an impactor in the extreme hyper-velocity collision regime where approach velocities are $>>$ 10 \si{kms^{-1}}}

\section{Approach}
\subsection{Launch Vehicle Options}
In this paper two trajectory options are presented, a direct path and one which entails a series of gravitational assists (GA). The direct option would need sufficient advance warning of the approach into the inner Solar System of 2I/Borisov and so would also require telescopes and detection systems which were not in place at the time of the discovery of 2I/Borisov. Nevertheless there is still value in analysing feasibility of a launch to this direct trajectory as it is relevant to potential future missions to ISOs yet to be discovered. At the time of writing this paper, the optimal launch date for the indirect trajectory is years away and so feasibility studies are of interest.\\

The standard escape trajectory performance parameter used to ascertain the capability of a launch vehicle is known as the 'Characteristic Energy', denoted $C_{3}$, and is defined as the square of the hyperbolic excess speed with respect to Earth, measured usually in units of \si{km^{2}s^{-2}}. The values of $C_{3}$ derived applicable to the two aforementioned trajectories are  $C_{3}$=30.1 \si{km^{2}s^{-2}} and $C_{3}$=47.8 \si{km^{2}s^{-2}} respectively for the direct and indirect trajectories in question. It is thus a matter of choosing launch vehicles which can deliver significantly large payload masses to these $C_{3}$ escape orbits.
Two of the most powerful launchers are adopted for the following research - first, the SpaceX Falcon Heavy which is currently in operation and the second, the NASA Space Launch System (SLS), will soon have its maiden flight. 
Parameters for these launch vehicles are given in Table \ref{table:launchers}. This table was constructed using data from \cite{Vardaxis2015P}. \hlreviewthree{The masses for this source are very marginally lower than those given by the primary source for such information which is http://elvperf.ksc.nasa.gov/ and indeed it is intended that this source shall be used for future investigations.} \hlreviewtwo{Note that the Falcon Heavy expendable is assumed.}\\


\begin{table}[]
\caption{Two Different Launcher Options for 2I/Borisov Missions (using \cite{Vardaxis2015P})}
\vspace{0.1 in}
\label{table:launchers}
\hspace*{-3.0cm}
\begin{tabular}{|c|c|c|c|c|}
\hline
Launch Vehicle  & \begin{tabular}[c]{@{}c@{}}Launch Site \\ for Eastwards\\  Launch Azimuth\end{tabular} & \begin{tabular}[c]{@{}c@{}}Geographic\\ Latitude\\ of Launch\\ Site\end{tabular} & \begin{tabular}[c]{@{}c@{}}Payload Mass \\ to \\ $C_{3}$=30.1 \si{km^{2}s^{-2}}\end{tabular} & \begin{tabular}[c]{@{}c@{}}Payload Mass\\  to \\ $C_{3}$=47.8 \si{km^{2}s^{-2}}\end{tabular} \\ \hline
SpaceX Falcon Heavy  & \begin{tabular}[c]{@{}c@{}}Kennedy Space Centre\\ LC-39A\end{tabular} & \ang{28.608389} & 8000 \si{kg} & 5000 \si{kg} \\ \hline
\begin{tabular}[c]{@{}c@{}}NASA Space Launch System\\ Block 1B\end{tabular}   & \begin{tabular}[c]{@{}c@{}}Kennedy Space Centre\\ LC-39B\end{tabular} & \ang{28.627222} & 23000 \si{kg} & 18000 \si{kg} \\ \hline
\end{tabular}
\end{table}

\subsection{Software, Assumptions and Limitations}
We used the Optimum Interplanetary Trajectory Software (OITS) developed by Adam Hibberd for calculating the trajectories \citep{OITS_info}.\\
OITS is based on a patched conic approximation, so within the sphere of influence of a celestial body only its gravitational attraction is taken into account. Outside of this sphere the spacecraft (s/c) is only influenced by the gravity of the sun. Also impulsive thrust is assumed, which is a valid approximation for high thrust propulsion systems such as the chemical rockets considered in this paper.\\
Note that by discounting perturbing forces, such as solar radiation pressure and gravitational perturbations due to non-homogeneous planetary mass distribution, this will compromise and reduce the payload mass results calculated later in Section \ref{sec:PPM}, as extra fuel will be needed for attitude and orbit control. The assumption  of impulsive thrust will have a similar effect, and efficiency losses due to this assumption are explored in \cite{Ferreira2017}\\  
\hlreviewtwo{It must be emphasised at this point that the software does NOT find a globally optimal combination of planetary encounters for a user-specified selection of $n$ planets. Rather it finds the optimal solution of encounter times at a user-specified sequence of $n$ planets in a user-specified order. Non-Linear Programming Software is exploited in order to determine the optimal $\Delta V$ which in turn is dependent on these $n$ times and must lie inside user-specified search intervals for these times. Thus global optimality cannot be guaranteed from this mode of analysis. However it is reiterated here that the purpose of this paper to find a \textbf{feasible solution}, not the precisely optimal one, and this is indeed achieved by the analysis which follows.}\\
Having designed the mission in terms of the sequence of celestial bodies to be visited, before a trajectory can be solved by the Non-Linear Programming Software, \hlreviewtwo{the user must constrain the search space.}\\
The trajectory is completely defined by the times of closest approach to each of the selected bodies in turn. Thus for the 'i-th' body, it has an associated time '$t_{i}$'. The time for the first body $i=1$ is the absolute Barycentric Dynamical Time whereas subsequent encounter times are the flight times to the $ith$ body from the preceding body $i-1$. Bounds on each of these parameters must be specified by the user, as well as various other optional constraints (such as total mission duration if desired). Given the body numbered $i$ in the sequence and its associated time, $t_{i}$  its position \textbf{$r_{i}$} (and for that matter its velocity \textbf{$v_{i}$}) can be derived, this obviously applies to the subsequent body $i+1$ also. If one assumes the connecting trajectory will only be influenced by the gravity of the sun, and ignoring perturbing forces, this problem can be formally stated as the 'Lambert Problem', which has two solutions, 'short way' and 'long way', corresponding to two different orbits. These can be solved by applying the Universal Variable Formulation, as elucidated in \citep{Bate1971}.\\ Thus if there are '$n$' such bodies along the trajectory, the number of possible trajectories for a given sequence of encounter times can be written as $2^{n-1}$. The permutation which minimises total $\Delta V$ is chosen. Total $\Delta V$ is defined here as the sum of all of the $\Delta V_{i}$ s at all the encounter bodies $i = 1,2,...n$. For a body not at the beginning or end of the sequence, the $\Delta V_{i}$ is assumed to be applied in the plane made by the approach velocity vector, \textbf{$VA_{i}$} and departure velocity vector \textbf{$VD_{i}$} with respect to the celestial body in question. Furthermore the $\Delta V_{i}$ is delivered at the periapsis point with respect to the body and is perpendicular to the radial vector from the body's centre. The $\Delta V_{i}$ at launch, i.e. $\Delta V_{1}$, is taken as the hyperbolic excess at Earth and the $\Delta V_{n}$ is zero, the latter because the missions here are all flybys of the target 2I/Borisov.\\

An 'Intermediate Point' is a point located in deep space which has zero mass and no associated velocity. However, its heliocentric radial distance is specified by the user and the heliocentric longitude and latitude are included with the encounter times as parameters for the optimizer. This feature allows OITS to model a Solar Oberth maneuver.\\

With these assumptions, a complete trajectory can be deduced as purely a function of the times $t_{i}, i=1,2...n$ and a total $\Delta V$ can be computed. 
 The resulting non-linear global optimization problem with inequality constraints is solved by the NOMAD solver \citep{LeDigabel2011}. OITS uses the ephemerides for all bodies calculated as a function of time using the NASA SPICE toolkit. The corresponding binary SPICE kernel file for 2I/Borisov was built on \hlreviewone{2020 February 3} by accessing the NASA HORIZONS telnet system.\\

\hlreviewtwo{The generally adopted minimum limit on periapsis altitude for each non-terminal encounter is 200 $\si{km}$. The optimized altitudes were generally much larger than this. The radius of Earth is taken as 6378 $\si{km}$ and that of Jupiter is 71492 $\si{km}$ (representing the 1 bar level).} \\
 
\hlreviewtwo{Note that a moon flyby can be used to leverage an interplanetary mission by reducing the launch $C_{3}$. However this option is not available in OITS, and indeed the intention of the paper is to address the principle feasibility of a mission to 2I/Borisov rather than to find precisely optimal solutions.}\\
 
\section{Results}
\subsection{Direct Transfer $\Delta V$ Contours}
\begin{figure}[h]
\hspace{-1.0cm}
\includegraphics[scale=0.42]{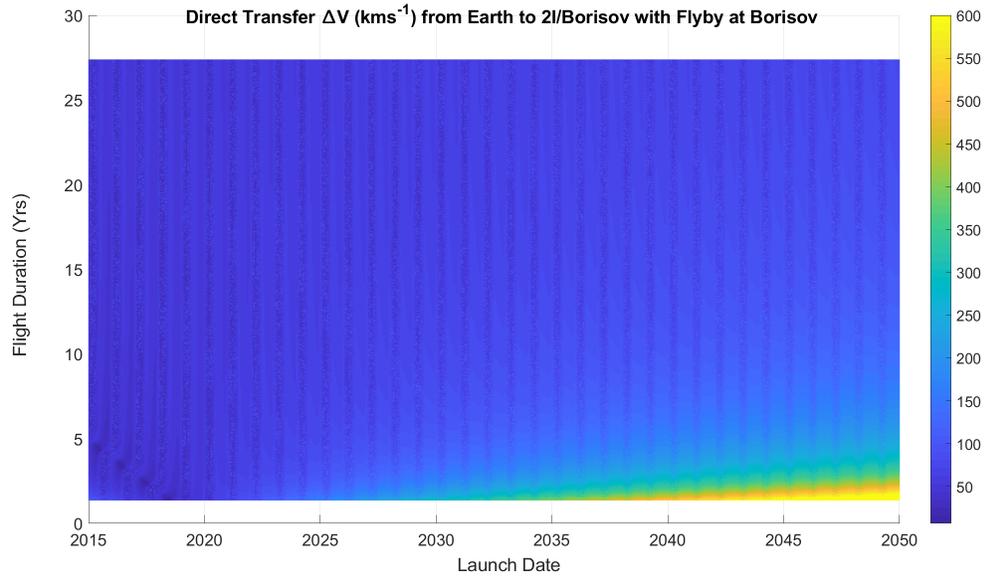}
\caption{$\Delta V$ color contours for trajectories to 2I/Borisov}
\label{fig:cont}
\end{figure}

Figure \ref{fig:cont} shows the $\Delta V$ contours for a direct transfer from Earth to 2I/Borisov with respect to flight durations up to 27 years (10,000 days) and launch dates from 2015 to 2050. The annual fluctuating repeating patterns in the direction of the launch date indicate the $\Delta V$ variation with respect to the Earth's position to the interstellar object. \hlreviewthree{Note that the $\Delta V$ in figure \ref{fig:cont} corresponds to hyperbolic excess speed at Earth (often designated $V_{\infty}$)}.It can be seen that for flight durations of more than 20 years, the total $\Delta V$ can be kept below 100 \si{km.s^{-1}}, even for launch dates in 2050. However, these $\Delta V$ values are still too high to achieve for existing chemical propulsion. \hlreviewtwo{If we filter figure \ref{fig:cont} and remove all trajectories with a $C_{3}$ $>$ 200 $\si{km^{2}s^{-2}}$ we get figure \ref{fig:cont_detail}. In this figure, the launch dates range from 2015 January to 2019 January, thus there are no direct trajectories which satisfy this $C_{3}$ requirement beyond 2019.} In the following section, we will present one minimal $\Delta V$ trajectory to 2I/Borisov, which would be feasible with existing chemical propulsion. 

\begin{figure}[h]
\hspace{-1.0cm}
\includegraphics[scale=0.39]{BORI_Detail.jpg}
\caption{$\Delta V$ detailed color contours for trajectory to 2I/Borisov}
\label{fig:cont_detail}
\end{figure}

\subsection{Optimal direct trajectory}
\begin{figure}[h]
\centering
\includegraphics[scale=0.37]{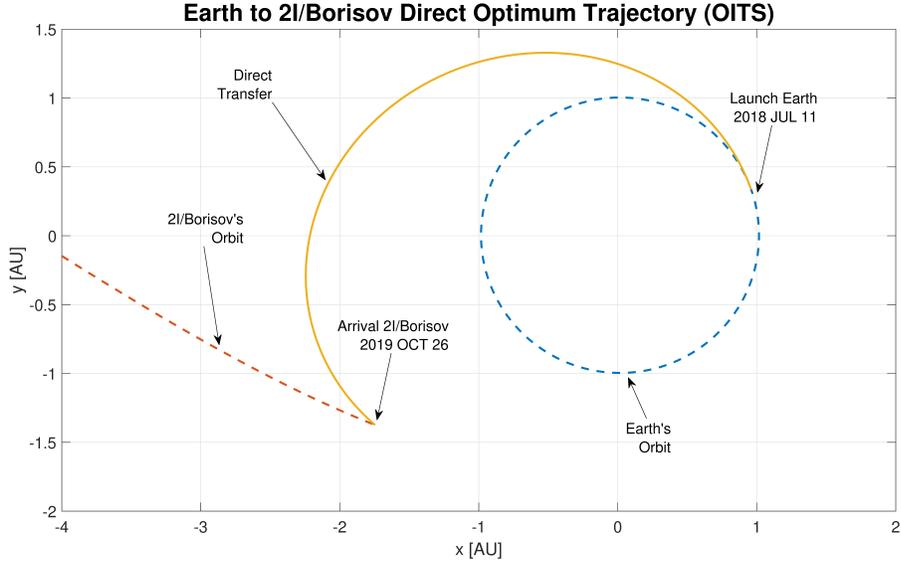}
\caption{Trajectory to 2I/Borisov}
\label{fig:bor}
\end{figure}

Figure \ref{fig:bor} shows the result for an optimal $\Delta V$ spacecraft trajectory (solid line) from Earth to recently discovered ISO 2I/Borisov.  The constraints on the launch date and flight duration were extremely large so as to ensure the solution was a global minimum. The plane of projection for Figure \ref{fig:bor} is that of the ecliptic, indeed this optimal transfer has a low inclination with respect to the ecliptic because it intercepts 2I/Borisov approximately when the ISO intersects the ecliptic plane. By so doing the s/c reaps maximum benefit from Earth's orbital velocity and the required $\Delta V$ is minimised. \hlreviewone{\textit{The launch date for this optimal trajectory was in 2018 July. Hence, we have missed the opportunity to send a spacecraft to 2I/Borisov directly at minimum $\Delta V$ (and indeed it was before the discovery date of 2I/Borisov in 2019 August).}}\hlreviewthree{After the discovery date in 2019 August, we find the next opportunity for an optimal direct mission lies on 2020 February 28, the corresponding $\Delta V$ depends on the flight duration selected, but if one selects a 450 day mission duration, we get the optimal value of 37.6 \si{kms^{-1}}.} \\
Table \ref{table:trajectory} shows key parameters for the trajectory. The spacecraft departs from Earth with a hyperbolic excess velocity of 5.5 \si{km.s^{-1}}. Arrival at 2I/Borisov is about one year later in \hlreviewone{2019 October}. Table \ref{table:directorb} provides the orbital elements in the NASA SPICE reference frame 'ECLIPJ2000', the Earth mean ecliptic and equinox of the epoch J2000. The maiden flight of the Falcon Heavy was in \hlreviewone{2018 February}, before this optimal launch date, and so it is pertinent to enquire whether such a trajectory could have been viable using a Falcon Heavy.\\
In assessing the viability of a particular launcher achieving injection directly into a specified escape orbit, the declination (denoted here as DEC) of the escape asymptote is a key parameter. A given Right Ascension can be attained by any launch system capable of Earth escape and depends on the relative alignment of the launch site with this (approximately) fixed direction. This alignment changes with time as the Earth spins on its axis and so it is a question of adjusting the launch time appropriately.\\
However the value of DEC is important and, in association with the launch site's latitude, is a key influence on the viability and characteristics of the launcher's ascent into the specified escape orbit. We find for the target escape orbit $DEC=\ang{8.544}$ (refer Table \ref{table:directorb}). \hlreviewone{Note that this value of DEC does not correspond to the declination of the outgoing radiant of 2I/Borisov, because the direct trajectory lies almost entirely in the ecliptic plane and the spacecraft would intercept 2I/Borisov at the descending node (approx.) of 2I/Borisov's orbit.} \hlreviewthree{It is re-emphasised at this point that DEC is \textit{relative to the equator NOT the ecliptic plane}} \\
\hlreviewthree{It should also be noted that generally, if the value of DEC is less than the latitude of the selected launch site then the escape asymptote can indeed be reached by the launch vehicle in question, often with a long coast arc in the final stage of the vehicle. Thus we find from Table \ref{table:launchers} that this mission is achievable by both of the assumed launchers, and indeed both launchers have a 2nd Stage restart capability.}

\begin{table*}
\centering
\caption{Values for minimal $\Delta V$ direct trajectory}
\vspace{0.1 in}
\label{table:trajectory}
\begin{tabular}{ |c|c|c|c|c|c|c| } 
\hline
 & Celestial body & Time & Arrival speed & Departure speed & $\Delta V$ \\
& & & \si{m.s^{-1}} & \si{m.s^{-1}} & \si{m.s^{-1}}\\
\hline
1 & Earth & \hlreviewone{2018 July 11} & 0 & 5485.3 & 5485.3\\ 
2 & 2I/Borisov & \hlreviewone{2019 October 26} & 33052.7 & 33052.7 & 0\\ 
\hline
\end{tabular}
\end{table*}

\begin{table}[]
\caption{Orbital Elements for Optimal Direct Trajectory to 2I/Borisov + RA \& DEC for Earth Escape Asymptote}
\vspace{0.1 in}
\hspace{-3.0cm}
\label{table:directorb}
\begin{tabular}{|c|c|c|c|c|c|c|c|c|c|c|c|}
\hline
\begin{tabular}[c]{@{}c@{}}Start\\ of \\ Arc\end{tabular} \begin{tabular}[c]{@{}c@{}}\end{tabular} & \begin{tabular}[c]{@{}c@{}}\hlreviewthree{True}\\ \hlreviewthree{Anomaly} \\  \hlreviewthree{($^{\circ}$)}\end{tabular} &\begin{tabular}[c]{@{}c@{}}End \\ of\\ Arc\end{tabular} & \begin{tabular}[c]{@{}c@{}}\hlreviewthree{True}\\ \hlreviewthree{Anomaly} \\  \hlreviewthree{($^{\circ}$)}\end{tabular} & \begin{tabular}[c]{@{}c@{}}\hlreviewone{$q$} Orbital\\ Parameter \\ (\hlreviewone{au})\end{tabular} & e & i ($^{\circ}$) & \hlreviewone{$\Omega$} ($^{\circ}$) & $\omega$ ($^{\circ}$) & DEC ($^{\circ}$) & \hlreviewone{RA} \\ \hline
\hlreviewone{2018 July 11} & \hlreviewthree{-0.23} & \hlreviewthree{2019 Oct 26} & \hlreviewthree{-161.33} & 1.0166 & 0.386 & 0.18 & -71.41 & 0.96 & 8.544 & \hlreviewone{01h09m00s} \\ \hline
\end{tabular}
\end{table}

With a $C_{3}$ of 30.1 \si{km^{2}.s^{-2}}, an existing \hlreviewone{expendable} Falcon Heavy launcher would \hlreviewone{have been} able to haul a payload of about eight \si{tonnes} to 2I/Borisov (refer Table \ref{table:launchers}) \hlreviewone{, had the discovery of 2I/Borisov been early enough}.\\
\hlreviewtwo{The phase angle (defined as the angle between the target's heliocentric radius and the approach velocity of the spacecraft) is $\ang{66.25}$, below the generally adopted maximum for such missions of $\ang{90.00}$}.\\

\subsection{Optimal Trajectories for later launch dates}
\label{sec:OTFLLD}
For later launch dates and a direct transfer, the $\Delta V$ increases to levels where no existing chemical propulsion system could deliver the required $\Delta V$. However an alternative approach is to use a Solar Oberth (SO) maneuver. \hlreviewthree{The specifics of such a maneuver are provided in detail in \cite{Friedman2014} in which this technique was explored in the context of an interstellar probe where large heliocentric speeds need to be generated in order to cover huge distances (on the order of $10^{2}$ au to $10^{3}$ au) very rapidly (20 or so years)}.\\
For a SO maneuver, the s/c is injected into a trajectory with a perihelion close to the sun, where the s/c applies a boost. The closer the boost is applied to the sun, the larger the performance benefit of the SO $\Delta V$ in terms of achieving the large heliocentric hyperbolic excess speed required to catch the target. Additional flyby maneuvers are used to bring the s/c on the initial heliocentric trajectory. We have previously shown that using a combination of planetary flybys and a SO maneuver, 1I/'Oumuamua can be reached with flight durations below 20 years and launch dates beyond 2030 \citep{Hein2019,Hibberd2020}.\\
Here, we shall use a combined Jupiter (J) powered flyby and SO maneuver to catch 2I/Borisov. This mission profile is given the abbreviation E-J-SO-2I. When the calculations are performed by OITS, there is some ambiguity as to whether the optimal launch year is 2030 or 2031, the former having the lower $\Delta V$, the latter having the lower in-flight time. Furthermore the optimal distance for the SO maneuver is reliant on several mission-dependent parameters. Practically speaking, the optimal combination of $\Delta V$ and flight time would depend on detailed study by the mission planners and there would be some trade-off between several other factors, like for instance the heat shield mass which would increase with reducing SO perihelion. \\

\begin{figure}[h]
\hspace{-1.0cm}
\includegraphics[scale=0.50]{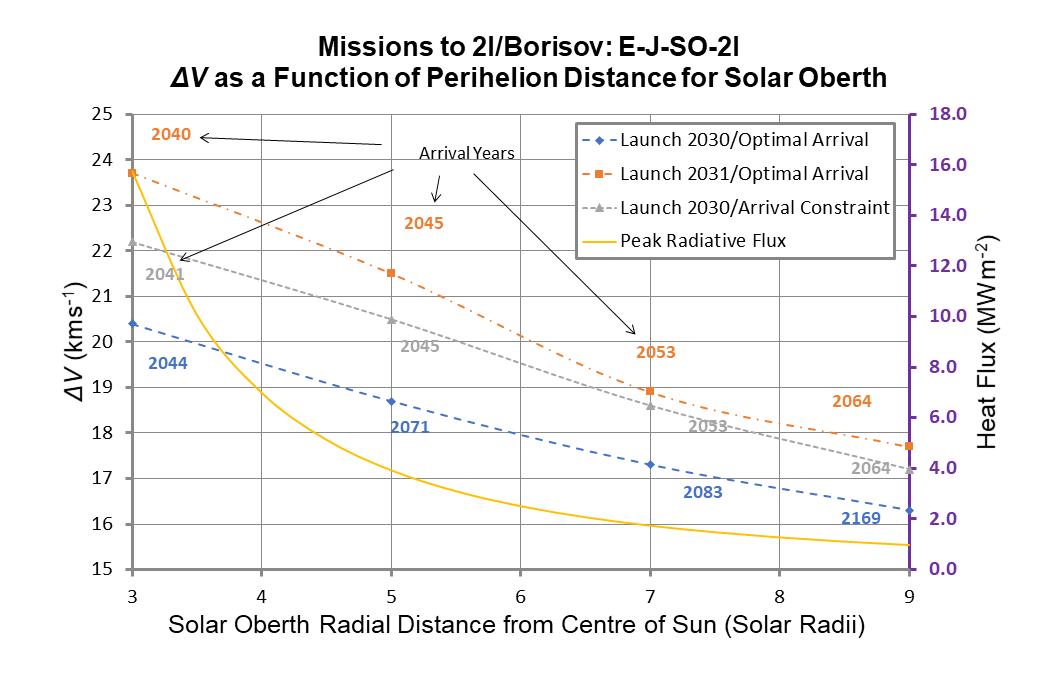}
\caption{Trajectory E-J-SO-2I for 3 different mission scenarios}
\label{fig:3-9SO}
\end{figure}

The situation is summarised by Figure \ref{fig:3-9SO} where 3 different mission scenarios are considered for a range of SO distances. Firstly the lowest blue dashed line shows what is achievable by an E-J-SO-2I scenario where the launch year is 2030 and the flight duration (and so arrival date) is optimal. The top red dotted and dashed line is the $\Delta V$ achievable by delaying the launch until 2031, with optimal flight duration. Finally the middle green dotted line is for a launch in 2030 but with the arrival date constrained to approximately satisfy the optimal arrival date achieved by the second scenario.\\
Observe that the optimal arrival year for a launch in 2030 (bottom dashed line) is always later than that achieved by a launch in 2031, with the discrepancy widening as the SO distance increases. Note however launching in 2031, the $\Delta V$s are considerably larger than launching in 2030. The middle green dashed line appears to be a good compromise, assuming a launch in 2030 arriving at 2I at an arrival date equivalent to that achieved by a 2031 launch but with a lower $\Delta V$.\\
As a case study from hereon we shall select the 2030 launch scenario, with a SO maneuver at 7 Solar Radii (SR) and an arrival year of around 2052/2053. This choice seems a reasonable compromise between $\Delta V$ and mission duration (24 years). \emph{Note that lower mission durations would be achievable but at the cost of higher $\Delta V$'s and correspondingly lower payload masses.}\\
Also present in Figure \ref{fig:3-9SO} is the peak solar flux variation with distance from sun. Observe at a distance of 7SR from the sun's centre, the solar flux is around 2.0 \si{MWm^{-2}} which is approximately twice that which the Solar Parker Probe (PSP) will experience at its closest perihelion. In the following it is assumed that an up-scaled version of the reinforced carbon-carbon composite heat shield used for PSP, can be applied to this mission, refer Section \ref{sec:PPM} for the calculations. \\
It has been demonstrated that preceding the flight from Earth to Jupiter by a $V_{\infty}$ Leveraging Maneuver has various benefits. \hlreviewthree{In this paper we remain consistent with previous research which utilise $V_{\infty}$ Leveraging Maneuvers \citep{Friedman2014,Hibberd2020}, in that we choose a resonance of 3 years. It should however be noted that a 2 year resonance is also possible, and has marginally lower $\Delta V$ (around 0.2 \si{kms^{-1}}lower). Amongst the benefits of such a maneuver are}:

\begin{enumerate}
    \item the hyperbolic excess speed (and so the $C_{3}$ value) required by the launch vehicle is reduced
    \item overall mission $\Delta V$ is reduced
    \item the mission launch window can usually be extended
    \end{enumerate}
    
For the $V_{\infty}$ Leveraging Maneuver, an approximately 3 year heliocentric ellipse is followed, beginning and ending at Earth, with a Deep Space Maneuver (DSM) in between after around 1.5 years and with an aphelion distance of 3.2 \hlreviewone{au}. This latter value is selected to ensure that the ellipse's major axis is equal to 1.0 \hlreviewone{au} + 3.2 \hlreviewone{au} = 4.2 \hlreviewone{au}, which corresponds to a time period of around 3 years, by Kepler's third law. After the return to Earth, a powered flyby is required to take the s/c to Jupiter whereupon the J-SO-2I leg of the trajectory continues as described above. The  $V_{\infty}$ Leveraging Maneuver is covered in detail in \cite{Sims1994AnalysisOV}. The overall mission trajectory can be abbreviated as E-DSM-E-J-7SR-2I. Table \ref{table:3SR} shows the results. 
\begin{figure}[h]
\centering
\includegraphics[scale=0.41]{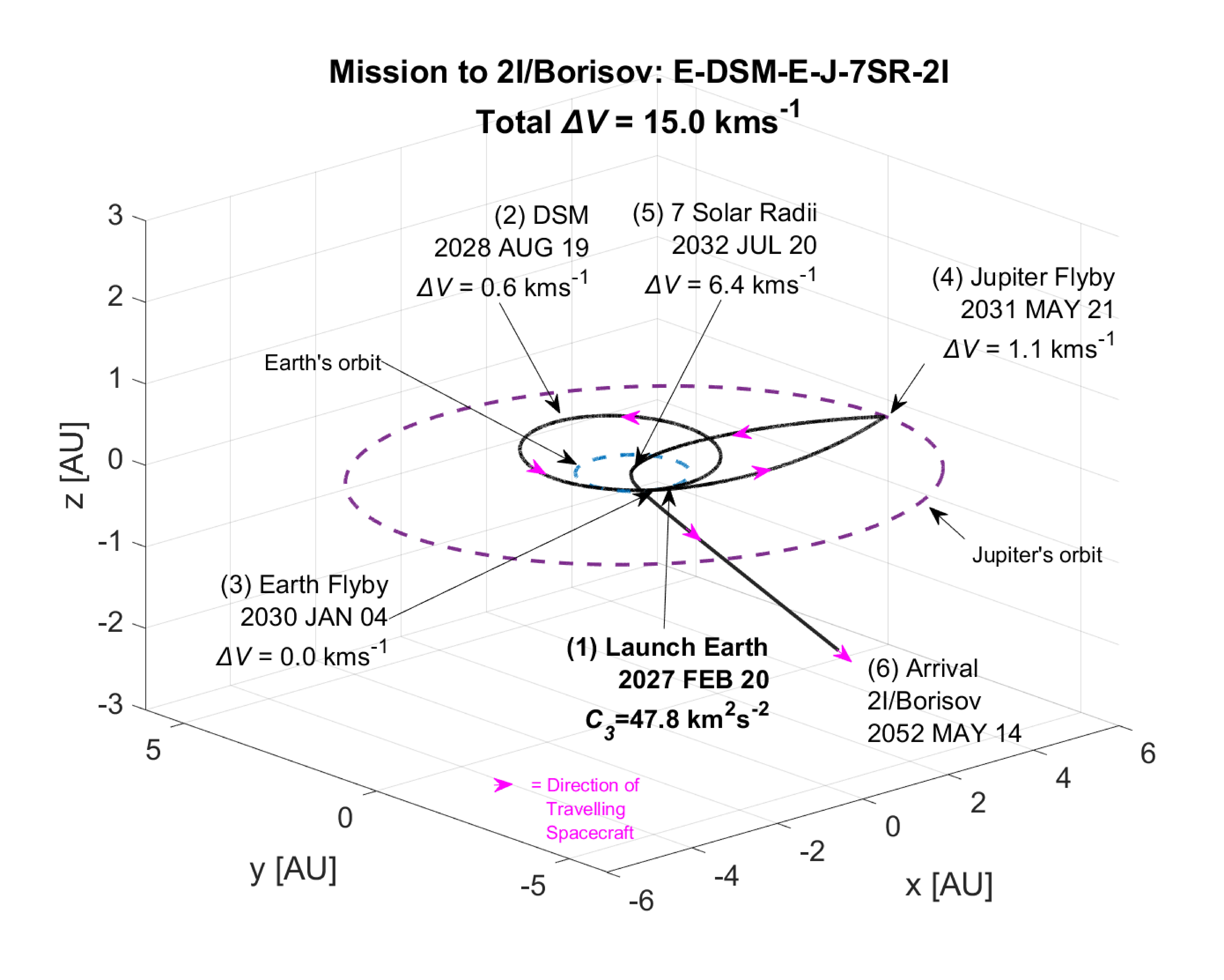}
\caption{Trajectory E-DSM-E-J-7SR-2I}
\label{fig:EDSMEJ7SR2I}
\end{figure}

\begin{table*}
\centering
\caption{Values for E-DSM-E-J-7SR-2I with powered Jupiter flyby}
\vspace{0.1 in}
\label{table:3SR}
\begin{tabular}{ |c|c|c|c|c|c|c| } 
\hline
 & Celestial & Time & Arrival speed & Departure speed  & $\Delta V$ & Cumulative \\
 & body &  & \si{km.s^{-1}} & \si{km.s^{-1}} & \si{km.s^{-1}} & $\Delta V$ \si{km.s^{-1}} \\
\hline
1 & Earth & \hlreviewone{2027 February 20} & 0 & 6.9137 & 6.9137 & 6.9137\\
2 & DSM & \hlreviewone{2028 August 19} & 11.4426 & 10.8855 & 0.557 & 7.4707\\
3 & Earth & \hlreviewone{2030 January 04} & 11.8532 & 11.8532 & 0 & 7.4707\\ 
4 & Jupiter & \hlreviewone{2031 May 21} & 14.0357 & 17.4855 & 1.1261 & 8.5968\\ 
5 & 7SR & \hlreviewone{2032 July 20} & 23.30546 & 239.4546 & 6.4377 & 15.0344\\ 
6 & 2I/Borisov & \hlreviewone{2052 May 14} & 20.8667 & 20.8667 & 0 & 15.0344\\ 
\hline
\end{tabular}
\end{table*}

\begin{table}[]
\centering
\caption{Orbital Elements for each leg of the Interplanetary Trajectory E-DSM-E-J-7SR-2I}
\vspace{0.1 in}
\label{table:indirectorb}
\hspace*{-3.0cm}
\begin{tabular}{|c|c|c|c|c|c|c|c|c|}
\hline
\begin{tabular}[c]{@{}c@{}}Start\\ of \\ Arc\end{tabular} & \begin{tabular}[c]{@{}c@{}}\hlreviewthree{True}\\ \hlreviewthree{Anomaly} \\  \hlreviewthree{($^{\circ}$)}\end{tabular} &\begin{tabular}[c]{@{}c@{}}End \\ of\\ Arc\end{tabular} & \begin{tabular}[c]{@{}c@{}}\hlreviewthree{True}\\ \hlreviewthree{Anomaly} \\  \hlreviewthree{($^{\circ}$)}\end{tabular} & \begin{tabular}[c]{@{}c@{}}\end{tabular} \begin{tabular}[c]{@{}c@{}}\hlreviewone{$q$}\\ Orbital\\ Parameter (\hlreviewone{au})\end{tabular} & $e$ & $i$ ($^{\circ}$) & \hlreviewone{$\Omega$} ($^{\circ}$) & $\omega$ ($^{\circ}$)\\ \hline
\hlreviewone{2027 February 20} & \hlreviewthree{0.25} & \hlreviewone{2028 August 19} & \hlreviewthree{179.05} & 0.9888 & 0.528 & 0 & 168.73 & 162.62 \\ \hline
\hlreviewone{2028 August 19} & \hlreviewthree{179.20} & \hlreviewone{2030 January 04} & \hlreviewthree{-47.02} & 0.8694 & 0.573 & 0 & 172.97 & 158.22 \\ \hline
\hlreviewone{2030 January 04} & \hlreviewthree{-17.70} & \hlreviewone{2031 May 21} & \hlreviewthree{139.80} & 0.9617 & 0.865 & 1.09 & 104.37 & -162.5 \\ \hline
\hlreviewone{2031 May 21} & \hlreviewthree{-174.18} & \hlreviewone{2032 July 20} & \hlreviewthree{-3.78} & 0.0325 & 0.993 & 109.48 & -98.17 & -5.38 \\ \hline
\hlreviewone{2032 July 20} & \hlreviewthree{-2.95} & \hlreviewone{2052 May 14} & \hlreviewthree{154.92} & 0.0325 & 1.104 & 109.49 & -98.17 & -5.88 \\ \hline
\end{tabular}
\end{table}

A data summary for each of the encounters along this E-DSM-E-J-7SR-2I trajectory is provided in Table \ref{table:indirectparam}. As explained for the direct case above, the most important parameter for launch trajectory considerations is the value of DEC for the home planet which from Table \ref{table:indirectparam} has a negative value of \ang{-20.979}. \hlreviewthree{This is of a magnitude less than that of the latitudes of the launch sites in Table \ref{table:launchers} and so the Earth escape trajectory is achievable by these launchers.} 

A secondary consideration along the interplanetary trajectory is the relative alignments of the s/c, the Earth and the Sun. \hlreviewthree{In particular two factors of importance are (a) whether there any periods where the spacecraft/Earth line-of-sight is obstructed by the sun and (b) whether there any periods where the spacecraft transits in front of the sun's disc. These factors affect the communications link between the Earth's Deep Space Network and the travelling s/c. Of particular relevance is the crucial period when the s/c is travelling close to the sun, including the burn at perihelion (at 7SR) known as the Solar Oberth maneuver. From Table \ref{table:indirectorb} one can find that the inclination of the trajectories firstly from Jupiter to 7SR and secondly from 7SR to 2I/Borisov are high at around \ang{109.5} with respect to the ecliptic. Indeed the s/c will be almost over the ecliptic North Pole with respect to the sun and so it is possible that no incidents of either (a) or (b) occur. To confirm this an analysis was conducted into periods of (a) or (b) along the E-DSM-E-J-7SR-2I trajectory. The assumption is of a spherical sun of $radius$ = 6963420 km. The results revealed that there are indeed \textbf{no periods of either obscuration or transit during this Solar Oberth Maneuver.}}

\hlreviewtwo{The phase angle between the spacecraft and 2I/Borisov for this mission is only $\ang{1.08}$ because the spacecraft approach velocity is near sun-radial with respect to Borisov}.\\
\begin{table}[]
\centering
\caption{Key Parameters for Each of the Encounters E-DSM-E-J-7SR-2I}
\vspace{0.1 in}
\hspace*{-3.0cm}
\label{table:indirectparam}
\begin{tabular}{|c|c|c|c|c|c|c|c|}
\hline
\textbf{Parameter} & \textbf{Units} & \textbf{Earth} & \textbf{DSM} & \textbf{Earth} & \textbf{Jupiter} & \textbf{7SR} & \textbf{2I/Borisov} \\
\textbf{} & \textbf{} & \textbf{} & \textbf{} & \textbf{} & \textbf{} & \textbf{} & \textbf{} \\ \hline
\textbf{Heliocentric Distance} & \hlreviewone{au} & 0.989 & 3.2 & 0.983 & 5.289 & 0.033 & 224.311 \\ \hline
\textbf{Heliocentric Longitude} & deg & 151.599 & -29.601 & 104.176 & -98.321 & 84.791 & -86.857 \\ \hline
\textbf{Heliocentric Latitude} & deg & -0.001 & 0.002 & -0.004 & 0.421 & 8.313 & -29.01 \\ \hline
\textbf{\begin{tabular}[c]{@{}c@{}}True Anomaly of\\ Object\end{tabular}} & deg & -106.631 & 180 & -107.536 & 85.045 & 180 & 103.37 \\ \hline
\textbf{\begin{tabular}[c]{@{}c@{}}Angle between Arrival and\\ Departure Asymptotes\end{tabular}} & deg & N/A & 0.006 & 34.822 & 107.827 & 0.169 & 0 \\ \hline
\textbf{Beta} & deg & N/A & N/A & 83.452 & 84.356 & N/A & N/A \\ \hline
\textbf{\begin{tabular}[c]{@{}c@{}}True Anomaly of Arrival\\ Asymptote\end{tabular}} & deg & N/A & N/A & 107.411 & 147.219 & N/A & N/A \\ \hline
\textbf{\begin{tabular}[c]{@{}c@{}}True Anomaly of Escape\\ Asymptote\end{tabular}} & deg & N/A & N/A & 107.411 & 140.608 & N/A & N/A \\ \hline
\textbf{Periapsis} & km & N/A & N/A & 6644 & 121806 & N/A & N/A \\ \hline
\textbf{Impact Parameter} & km & N/A & N/A & 8768 & 311544 & N/A & N/A \\ \hline
\textbf{Miss Distance} & km & N/A & N/A & 9047 & 414120 & N/A & N/A \\ \hline
\textbf{Arrival Hyperbolic Excess} & \si{kms^{-1}} & N/A & 11.4426 & 11.8532 & 14.0357 & 23.30546 & 20.8667 \\ \hline
\textbf{Arrival Right Ascension} & deg & N/A & 57.131 & 257.514 & 237.542 & 51.971 & 272.997 \\ \hline
\textbf{Arrival Declination} & deg & N/A & 20.004 & -22.938 & -19.094 & -55.923 & -52.964 \\ \hline
\textbf{Departure Hyperbolic Excess} & \si{kms^{-1}} & 6.9137 & 10.8855 & 11.8532 & 17.4855 & 239.4546 & 20.8667 \\ \hline
\textbf{Departure Right Ascension} & deg & 242.243 & 57.125 & 222.551 & 135.498 & 52.101 & 272.997 \\ \hline
\textbf{Departure Declination} & deg & -20.979 & 20.004 & -12.431 & 22.192 & -55.771 & -52.964 \\ \hline
\end{tabular}
\end{table}

\subsection{Potential Payload Masses}
\label{sec:PPM}
It is now a question of translating the $\Delta V$s from Table \ref{table:3SR} into payload mass to 2I/Borisov.\\
From Table \ref{table:3SR} we have a launch $C_{3}$ of 47.8 \si{km^{2}.s^{-2}} and an overall $\Delta V$ of 15.0 \si{km.s^{-1}} Observe that the 7SR SO boost requires a $\Delta V$ of 6.4 \si{km.s^{-1}} and the remaining maneuvers come to a magnitude of 1.7 \si{km.s^{-1}}. The SO $\Delta V$ is not achievable by a single solid rocket booster, but two staged boosters are quite capable of delivering this kick.\\
A staged pair of boosters with optimal mass ratio (i.e. optimal in terms of achieving the highest payload mass) may be calculated. For this calculation, we assume that the $\Delta V$ is 6.4 \si{km.s^{-1}}, that both boosters have an exhaust velocity of 2.9 \si{km.s^{-1}}, i.e. a specific impulse ($I_{sp}$) of 296 \si{s} (equivalent to that of a STAR Solid Rocket Engine), and that the ratio of dry mass to wet mass for both stages (excluding payload) is $p=0.058$. We further select a value of total s/c mass prior to this SO (including payload) of 10000 \si{kg}. \hlreviewthree{The reason why this value is chosen will be elucidated further down.} Table \ref{table:SOMSLS} provides the results and indicates a payload mass of 841 \si{kg}. Subtracting the mass of the heat shield, this gives an eventual weight of 765 \si{kg}.\\
This heat shield is taken as proportional to the squared cube-root of the overall payload mass (refer also Section \ref{sec:OTFLLD}), scaling up or down from the  Parker Solar Probe overall mass/heat shield mass accordingly \hlreviewthree{where the PSP shield mass is taken to be 66kg}). Thus the mass of the reinforced carbon-carbon composite heat shield is assumed proportional to the surface area of the s/c.\\ 
A SLS Block 1B with a Delta Cryogenic Second Stage (DCSS) would result in an 18000 \si{kg} payload to an escape trajectory with $C_{3}$ of 47.8 \si{km^2.s^{-2}} Furthermore With the performance increase in the Block 1 since its early conception, a value approaching this would also be achievable using a SLS Block 1.\\
We have dealt with the propulsion system necessary for the SO, there are however two burns prior to this at:
\begin{enumerate}
    \item The Deep Space Maneuver following 1.5 years after Earth launch
    \item Encounter with Jupiter
    \end{enumerate}
The magnitude of these can be found in Table \ref{table:3SR} and sum to approximately 1.7 \si{kms^{-1}}. If a throttleable Liquid Propellant Engine (assuming $MMH/N_2O_4$ fuel, $I_{sp}$=341 \si{s}, $p=0.1$, \hlreviewtwo{$m_{stage}$} = 18000 \si{kg} - 10000 \si{kg} = 8000 \si{kg}) is used to achieve these burns then it would be able to deliver a 10000 \si{kg} payload as required. \hlreviewthree{Thus we see the reason why 10000 \si{kg} was chosen above, it is because a throttleable Liquid Propellant stage of 8000 \si{kg} (with the aforementioned characteristics and using the rocket equation) has the capability of accelerating 10000 \si{kg} to the desired $\Delta V$ of 1.7 \si{kms^{-1}}}.\\
Falcon Heavy would be able to deliver 5000 \si{kg} to a $C_{3}$ of 47.8 \si{km^{2}.s^{-2}}. For calculating the payload mass, we use the same technique as used for the SLS, but this time with a s/c mass prior to the SO (including payload) of 2788 \si{kg}. From Table \ref{table:SOMFH}, this can deliver a total of 202 \si{kg} after the heat shield mass has been subtracted. The value of 2788 \si{kg} is chosen because a scaled down version of the Liquid Propellant Engine used in the calculation of the SLS can be used for Falcon Heavy but with \hlreviewtwo{$m_{stage}$} = 5000 \si{kg} - 2788 \si{kg} = 2212 \si{kg}.\\

\hlreviewthree{Comparing the mass for SLS of 765 \si{kg} and for Falcon Heavy of 202 \si{kg}, with the mass, 544 \si{kg}, of the interstellar mission probe outlined in \cite{Friedman2014}, we see that there would appear to be plenty of scope for SLS, but the Falcon Heavy payload mass is far less than the mass budget calculated from this source.}

\begin{table*}
\centering
\caption{Requirements for a SO Maneuver - Space Launch System}
\vspace{0.1 in}
\label{table:SOMSLS}
\begin{tabular}{ |c|c|c|c|c|c| }
\hline
 & Total &	Mass of Stage  & Mass &  Mass After & $\Delta V$ \\
& Initial &	Excluding  & Achieved after & Dead Weight & (\si{km.s^{-1}}) \\
& Mass (\si{kg}) & payload (\si{kg})  & Stage Burn (\si{kg}) &  Jettison (\si{kg}) & \\
\hline
First Stage Data & 10000  & 6605  & 3778  & 3395 & 2.8 \\
\hline
Second Stage Data & 3395 & 2554 & 988 & 841 & 3.6 \\
\hline
Mass of Heat Shield & & & & 76 & - \\
\hline
Total & & & & \textbf{\emph{765}} & 6.4 \\
\hline
\end{tabular}
\end{table*}

\begin{table*}
\centering
\caption{Requirements for a SO Maneuver- Falcon Heavy}
\vspace{0.1 in}
\label{table:SOMFH}
\begin{tabular}{ |c|c|c|c|c|c| }
\hline
 & Total &	Mass of Stage  & Mass &  Mass After & $\Delta V$ \\
& Initial &	Excluding  & Achieved after & Dead Weight & (\si{km.s^{-1}}) \\
& Mass (\si{kg}) & payload (\si{kg})  & Stage Burn (\si{kg}) &  Jettison (\si{kg}) & \\
\hline
First Stage Data & 2788  & 1842  & 1053  & 946 & 2.8 \\
\hline
Second Stage Data & 946 & 712 & 276 & 234 & 3.6 \\
\hline
Mass of Heat Shield & & & & 32 & - \\
\hline
Total & & & & \textbf{\emph{202}} & 6.4 \\
\hline
\end{tabular}
\end{table*}

\section{Discussion}
It would appear that although 2I/Borisov presents a considerable challenge for an intercept mission, in terms of its high hyperbolic excess speed (around 32 \si{km.s^{-1}}) and high orbital inclination with respect to the ecliptic (44 \si{\degree}), nevertheless a mission would be achievable with chemical propulsion technology in the form of the currently operating SpaceX Falcon Heavy lift vehicle or the forthcoming NASA Space Launch System and \hlreviewthree{scaled versions of existing technologies}.  There is some debate as to the discovery rate of ISOs when the Vera C. Rubin observatory (also LSST) comes into operation in the near future, \cite{Siraj_2020} have determined it would be around one or two per month. Presumably future ISOs will present similar difficulties and the evidence on the basis of the two known ISOs which have been studied here indicates these objects, despite stretching the limits of current technology, are indeed reachable. \\
Note that launch date optima for missions to 2I/Borisov will repeat on a cycle of around 12 years (which is roughly Jupiter's orbital period), thus although the mission studied here required a launch in 2027, other opportunities will also arise in 2039, 2051,...etc. The drawback to delaying a mission in this way is that 2I/Borisov will have receded from the sun significantly and so the catch-up time will be necessarily extended. Assuming a s/c hyperbolic excess of 53 \si{km.s^{-1}}, and using the value of 2I/Borisov's excess of 32 \si{km.s^{-1}} , we can deduce that the travel time extends by around 18.5 years per 12 year delay in launch. Furthermore, doubtless R\&D will continue in the intervening years and more advanced propulsion schemes will be developed allowing faster trajectories with more massive payloads.\\
It should be mentioned here that the Comet-Interceptor mission \citep{2020LPI....51.2938J}, in which the s/c lies in wait at the Sun-Earth L2 point for the detection of a suitable long period comet and then is dispatched along an appropriate trajectory to intercept and study it, may alternatively be deployed to an ISO. This architecture is not currently appropriate for ISOs of particular interest because of unusual properties (1I/'Oumuamua) and which may have traveled past their perihelia because they were discovered too late.

\section{Conclusion}
This paper assesses the technical feasibility of a mission to 2I/Borisov, using \hlreviewthree{scaled versions of} existing technologies. We apply the Optimum Interplanetary Trajectory Software (OITS) tool to generate direct and indirect trajectories to 2I/Borisov. The direct trajectory with a minimal $\Delta V$ is identified as the one with a launch date in \hlreviewone{2018 July}. For this trajectory, a Falcon Heavy launcher could have hauled an 8 ton spacecraft to 2I/Borisov. For indirect trajectories with a later launch date, results for a combined powered Jupiter flyby with a SO maneuver are presented. For a launch in 2027, we could reach 2I/Borisov in 2052, using the Space Launch System, up-scaled Parker probe heat shield technology, and solid propulsion engines, a 765 kg spacecraft could be sent to 2I/Borisov. A Falcon Heavy can deliver 202 kg to 2I/Borisov. Arrival times sooner than 2052 can be achieved, but at the cost of increased $\Delta V$ and consequently lower payload masses. With the Vera C. Rubin observatory (also known as LSST) coming into operation in the near future, it is expected more ISOs will be discovered. The feasibility of a mission to both 1I/'Oumuamua and 2I/Borisov using \hlreviewthree{scaled versions of} existing technologies indicates that missions to future interstellar objects are likely to be feasible as well.

\bibliographystyle{model5-names}
\biboptions{authoryear}
\bibliography{library,library_Adam_Hibberd,Hein_ISO_Modified_by_Adam_Hibberd}

\end{document}